\documentclass[aps,prd,twocolumn,groupedaddress,nofootinbib]{revtex4-1}

\pdfoutput=1

\usepackage[english]{babel}
\usepackage[latin1]{inputenc}
\usepackage{graphicx}
\usepackage{amsmath}
\usepackage{color}

\begin{document}

\title{Fractal universe and cosmic acceleration \\
in a Lema\^{i}tre-Tolman-Bondi scenario}
%
%
\author{L. Cosmai} \email{leonardo.cosmai@ba.infn.it}
\affiliation{Istituto Nazionale di Fisica Nucleare, Sezione di Bari, via G. Amendola 173, 70126, Bari, Italy}

\author{G. Fanizza} \email{giuseppe.fanizza@pi.infn.it}
\affiliation{INFN - Sezione di Pisa, Largo B. Pontecorvo 3, I-56127 Pisa, Italy}
\affiliation{Center for Theoretical Astrophysics and Cosmology,
Institute for Computational Science, University of Z\"urich, Winterthurerstrasse 190, CH-8057, Z\"urich, Switzerland}

\author{F. Sylos Labini}\email{sylos@centrofermi.it}
\affiliation{Museo  Storico  della  Fisica   e  Centro  Studi  e
          Ricerche Enrico  Fermi, I-00184 Rome, Italy} \affiliation{Istituto dei
          Sistemi Complessi, Consiglio Nazionale delle Ricerche, I-00185
          Roma, } \affiliation{Istituto  Nazionale  Fisica  Nucleare,
  Dipartimento  di  Fisica, Universit\`a  ``Sapienza'',  I-00185
          Roma,  Italy}

\author{L. Pietronero}
\email{luciano.pietronero@roma1.infn.it}
\affiliation{ Dipartimento  di  Fisica, Universit\`a  ``Sapienza'',  I-00185
          Roma,  Italy}
\affiliation{Istituto dei
          Sistemi Complessi, Consiglio Nazionale delle Ricerche, I-00185
          Roma}

\author{L. Tedesco}
\email{luigi.tedesco@ba.infn.it}
\affiliation{Istituto Nazionale di Fisica Nucleare, Sezione di Bari, Bari, Italy}
\affiliation{Dipartimento di Fisica, Universit\`a di Bari, Via G. Amendola 173, 70126, Bari, Italy}

\date{\today}

\begin{abstract}In this paper we attempt to answer to the question: { can cosmic
  acceleration of the Universe have a fractal solution?}  We give an
exact solution of a Lemaitre-Tolman-Bondi (LTB) Universe based on 
  the assumption that such a smooth metric is able to describe, on
  average, a fractal distribution of matter. While the LTB model has a
  center, we speculate that, when the fractal dimension is not
    very different from the space dimension, this metric applies to
  any point of the fractal structure when chosen as center so that, on
  average, there is not any special point or direction. We examine
the observed magnitude-redshift relation of type Ia supernovae (SNe
Ia), showing that the apparent acceleration of the cosmic expansion
can be explained as a consequence of the fractal distribution of
matter when the corresponding space-time metric is modeled as a smooth
LTB one  and if the fractal dimension on scales of a few hundreds
  Mpc is $D=2.9 \pm 0.02$.
\end{abstract}

\pacs{}

\maketitle

The so-called concordance model of the universe is based on {\it
  three} fundamental assumptions.  The first is that the dynamics of
space-time is determined by Einstein's field equations. The second is
 the {\it generalization}  of the Copernican Principle
  \cite{bondi,ferreira}: all observers are equivalent and there are no
  special points and directions. The third is that matter
distribution is spatially homogeneous, i.e. characterized by a
spatially constant density $\rho(r,t)=\rho(t)$.  The
Friedmann-Lemaitre-Robertson-Walker (FLRW) model is derived under these three
assumptions and it describes the geometry of the universe in terms of
a single function, the scale factor, which obeys to the Friedmann
equation \cite{Weinberg}.   In this framework one  assumes
  exact translational and rotational invariance. When density
  fluctuations are introduced, their possible compatibility with the
  above framework depends on fluctuations statistical properties.

 If matter distribution is a realization of a {\it stationary}
   stochastic point process \cite{book}, it is statistically
   homogeneous and isotropic thus still satisfying the Copernican
   requirement of the absence of special points or directions.
   However it can, or cannot be a spatially homogeneous stationary
   stochastic process: only in the latter case matter distribution
   satisfies the special and stronger case of the Copernican Principle
   described by the Cosmological Principle.  Indeed, isotropy around each
   point together with the hypothesis that the matter distribution is
   a smooth function of position i.e., that this is analytical,
   implies spatial homogeneity \cite{{Trumper},{Straumann74}}.  This is no longer
   the case for a non-analytic structure (i.e., not smooth), for which
   the obstacle to applying the FLRW solutions has in fact solely to do
   with the lack of spatial homogeneity \cite{Anderson}.

Recently the attempts to construct cosmological models including
spatial inhomogeneities have experienced a renewed interest in
connection with both the detection of a complex network of galaxy
  clusters, filaments and void \cite{sdss} and the evidences for a
  speeding up expansion of the universe as shown by the supernovae
  (SN) observations \cite{Riess,Perlmutter}.  Indeed, the deduction
of the existence of dark energy is based on the assumption that the
universe has a FLRW geometry. The underlying idea of inhomogeneous
models is to interpret the acceleration derived from the SN
observations as an apparent effect that arises in a too simplified
solution of Einstein's field equations, i.e. those derived with a
density and pressure that are constant is space and depend only on
time. It is possible to distinguish \cite{Anderson_Coley} between
  two main approaches to an inhomogeneous universe: (i) models that
  consider the spatial averaging of inhomogeneities and (ii) models
  placing the observer in a special point of the local universe.

It has been demonstrated that the spatial averaging of inhomogeneities
gives rise to effective terms (named backreaction), in addition to the
standard FLRW energy sources, that can play the role of dark energy on
large scales  \cite {buc}. Dark energy can be thus considered as an artifact due
to the impact of inhomogeneities. There is an ongoing debate about
whether an inhomogeneous model can thus evolve, on average, like the
homogeneous FLRW solution in agreement with observations
\cite{Bonnor,Ellis_2008,Buchert_2008,Buch,Kolb_Marra_Matarrese,Larena_etal}

Models of type (ii) instead use a very ad-hoc behavior of the spatial
density implying a breakdown of the Copernican assumption on the
Hubble scale: in particular it was assumed that we are near the center
of spherically symmetric low density hole
\cite{Celerier,Anderson_Coley}. In this context the simplest GR models
are the spherically symmetric Lemaitre-Tolman-Bondi (LTB) solutions
with a central observer which clearly represent drastic simplification
of the problem. LTB models without dark energy can fit supernovae,
explaining the apparent acceleration of the universe by a Gpc-scale
void around us \cite{Celerier_2000}, thus requiring fine-tuned
  and ad-hoc assumptions on the properties of matter distribution.
\\
Another way to consider the effect of the inhomogeneities on the observations
consists of studying the effect of randomly distributed inhomogeneous patches
in a given background on the propagation of photon \cite{Kantowsi}. These attempts
are known as Swiss cheese models. In this approach, light rays travel along
a series of inhomogeneous patches which are usually
modeled with radially inhomogeneous LTB regions \cite{Marra1,Marra2} or
with Szekeres metric \cite{Bolejko,Peel,Koksbang}.
\\
An interesting attempt, concerning
a relativistic model for the observed large scale
inhomogeneities
was proposed by  \cite{Ribeiro:2008rs}: in this framework one makes
 the hypothesis that the large scale structures
can be described as being a self similar fractal system.
\\  
In this paper we adopt a different approach that consists in modeling
the spatial inhomogeneities in a way that is more close to what we can
learn from observations of galaxy structures.  Indeed, the statistical
analysis of galaxy three dimensional surveys have shown that galaxy
distribution is characterized by power-law correlations in the local
universe \cite{SLMP_1998,Hogg_2005,FSL_2011}. More specifically it was
found that the average conditional density decays as $\langle n(r) \rangle_p
\sim r^{-\gamma}$ where $\gamma=0.9 \pm 0.1$ for $r \in [0.1,20]$
Mpc/h and $\gamma=0.2 \pm 0.1$ for $r \in [20,100]$ Mpc/h
\cite{Antal}. Whether or not on scales $r> 100$ Mpc/h correlations
decay and the distribution crossovers to uniformity, is still matter
of considerable debate \cite{FSL_2011}. 

The power-law behavior of the conditional density can be interpreted
as galaxy distribution having fractal properties at small scales.  A
fractal is a non-analytical point distribution:  $\langle n(r)
\rangle$ decays (only) on average as a power law.  From the
  $i^{th}$ the (conditional) density decays as $n_i(r) \sim f_i(r) \cdot 
  r^{-\gamma}$ where the correction to scaling $f_i(r)$ is such that
  $\sigma_p(r) = \langle n_i(r)^2 \rangle - \langle n(r) \rangle^2_p =
  \langle (f(r) -1)^2 \rangle \approx$ const. \cite{book}. If
  $\sigma_p(r) < 1$, as for the real galaxy structures \cite{Antal},
  we can can approximate the discrete matter source field as  
\begin{equation}
\label{ddensity} 
\rho_d(r) = \sum_i m_i \delta^D(\bar{r}-\bar{r}_i) 
\approx \langle n(r) \rangle \;. 
\end{equation}
This situation allows us to use a smooth LTB metric for describing
the spatial decay of the density
without assuming the existence of a special position in the Universe. Indeed, 
in LTB models isotropy is valid only for the privileged observer that makes measurements from the centre of coordinate system. Any other observer in a LTB Universe far from the centre will experience a dipolar anisotropy  \cite{alnes_2006,alnes_2007,fanizza_2015}  and thus the 
 Cosmological Principle is not valid in such a framework. 
 On the other hand, a fractal distribution has the fundamental property that the density is seen to decay with the same power law for all the observers: as mentioned above
 it can be seen a stationary point process, i.e. a statistically homogeneous 
 and isotropic distribution.
 To reconcile these two different properties of the metric of and of matter 
 distribution, we assume to live in a local
over-density and that any other observer, placed in any other galaxy,
sees the same decaying radial density as us: this condition 
is {\it approximately} satisfied if $\sigma_p(r) <1$ and thus Eq.\ref{ddensity} holds.
  This situation has a
clear advantage with respect of assuming a single large-scale
Gpc under-density.  Indeed, while the LTB model has a center, we
speculate that this metric applies to any point of the fractal
structure when chosen as center so that, on average, there is not any
special point or direction.

    We are not able to quantify the
perturbations neglected by making this assumption but we can assume
that, as long as spatial fluctuations around the average behavior
remain limited, i.e.  $\sigma_p(r) < 1$, this model provide  a
reasonable description of the local metric of a fractal object.
In this situation the local expansion rate around us would be smaller
than the average expansion rate in the background: light-rays
propagating from distance sources to us (or to any observer located in
a local over-density) would therefore feel a {\it decelerated}
expansion rate along their path: this is what we are going to show. 
\\
\\
As discussed below we do not need to assume that the fractal behavior
extends up to an arbitrarily large scale:  rather we show that 
 it is sufficient that for a moderate value
of the homogeneity scale $l_0$ , beyond which $\langle n(r) \rangle_p
\approx const.$, the modification of the magnitude-redshift relation
due to the inhomogeneous and power-law behavior of the (conditional)
density provides a best fit to the SN data without the
need of introducing dark energy.

Let us describe an inhomogeneous universe within the framework of the
isotropic and inhomogeneous LTB metric in polar coordinates
$x^{\mu}=(t,r,\theta,\phi)$
\small
\begin{equation}
ds^2=-dt^2+\frac{A'(t,r)^2}{1-k(r)}dr^2+A^2(t,r) \left[ d\theta^2+\sin^2\theta d\phi^2 \right]
\label{eq:line_element}
\end{equation}
\normalsize
where $A(t,r)$ is the radial inhomogeneous scale factor and $k(r)$ is
the inhomogeneous spatial curvature. 
 Therefore the two independent Einstein's equations
which describe the model are:
\begin{align}
\label{eq:EinsteinEqs}
\frac{\dot A^2+k}{A^2}+ 
\frac{2\dot A \dot A'+k'( r)}{A A'}&=8\pi G \langle n(r) \rangle   \\
\label{eqA}
\frac{\dot A^2+2A\ddot A+k}{A^2}&=0
\end{align}
where $\dot{}\equiv\partial_t$ and $'\equiv\partial_r$. By integrating the eq.~(\ref{eqA}), we get:
\begin{equation}
\label{eq:firstIntegration}
\left( \frac{\dot A}{A} \right)^2=-\frac{k( r)}{A^2}+\frac{\alpha(r)}{A^3}
\end{equation}
which, for $k=0$ can be exactly integrated with solution:
\begin{equation}
\label{eq:solution}
A(t,r)=A_0( r)\left[ 1+\frac{3}{2} \sqrt{\frac{\alpha(r)}{A_0^3( r)}}\,t \right]^{2/3} \, ,
\end{equation}
where $A_0( r)$ and $\alpha( r)$ are two free functions due to the double integration in $t$.
Here we adopt the simplified choice $k=0$ because it is supported by the value of spatial curvature on cosmological scales is highly constrained by the CMB best-fit parameters to be very small in the past and negligible today \cite{Planck}. This assumptions implies that the inhomogeneous behavior on large scales is not due to curvature perturbations.
Let us now define the Hubble function $H(t,r)=\dot A/A$ and the
comoving mass  
\begin{equation}
M(t,r)=\int_{S_P^3(r)} \langle n(r) \rangle  A' A^2 4 \pi r^2 d r
\end{equation}
where $S_P^3(r)$ is the 3D sphere with radius $r$ and
centered in the observer position $P$. By inserting the solution
eq.\eqref{eq:solution} in eq.\eqref{eq:EinsteinEqs} and by 
integrating over the sphere; we get:
\begin{equation}
\alpha(r)=2 G M(r) 
=2 G \int_{S_P^3( r)} \langle n(r) \rangle  A'(t,r)A^2(t,r) 4 \pi r^2 d r 
\end{equation}
so that $M(r)$ only depends by $r$. 
From
Eq.\eqref{eq:firstIntegration}  we find
\begin{equation}
2GM( r)=A^3_0( r)H^2_0( r) \;, 
\end{equation}
where
$H_0( r)\equiv H(0,r)$. Therefore we get that the solution is
$A(t,r)=A_0( r)\left[ 1+\frac{3}{2}H_0(r)\,t \right]^{2/3}$ and the
Hubble function at the present time is intimately related to the mass
distribution by:
\begin{equation}
H_0(r)=\sqrt{\frac{2GM(r)}{A_0^3( r)}}.
\label{eq:constraint}
\end{equation}
Eq. \eqref{eq:constraint} naturally appears from the solution of the Einstein equations and is completely general (modulo the absence of spatial curvature) and it basically relates the inhomogeneous Hubble flow the mass content.
For a pure fractal we have $M(r) \sim r^{D}$ where $D=3-\gamma < 3$ is
the fractal dimension \cite{book}. The last free function $A_0(r)$ can be 
chosen  by exploiting the residual freedom that we have in redefining the radial coordinate $r$ in the Eq. \eqref{eq:line_element}. In fact, thanks to this we can always specify the function $A(t,r)$ at a given time $t_*$. Thanks to this we can choice $A_0(r)=r$ . Being this just due to a redefinition of coordinates that leaves us within the LTB metric, this choice cannot affect any observables.
We can write the mass $M(r)$ as
\begin{equation}
M( r)=\Phi\,r^D
\label{eq:MatterContent}
\end{equation}
where $\Phi$ is the amplitude of the
fractal distribution, related to the average distance between
  nearest galaxies \cite{book}. Therefore, we can rewrite the
Hubble function in a more intuitive way as:
\begin{equation}
H_0( r)=B\,r^\frac{D-3}{2}
\end{equation}
where $B\equiv\sqrt{2G\,\Phi}$. From the mathematical point of view,
let us notice that our expression for $H_0( r)$ diverges when
$r\rightarrow 0$. Nevertheless, we will argue later that this bad
behavior can be easily fixed by requiring an appropriate lower cutoff.

\begin{figure}[h!]
\centering
\includegraphics[width=10.5cm]{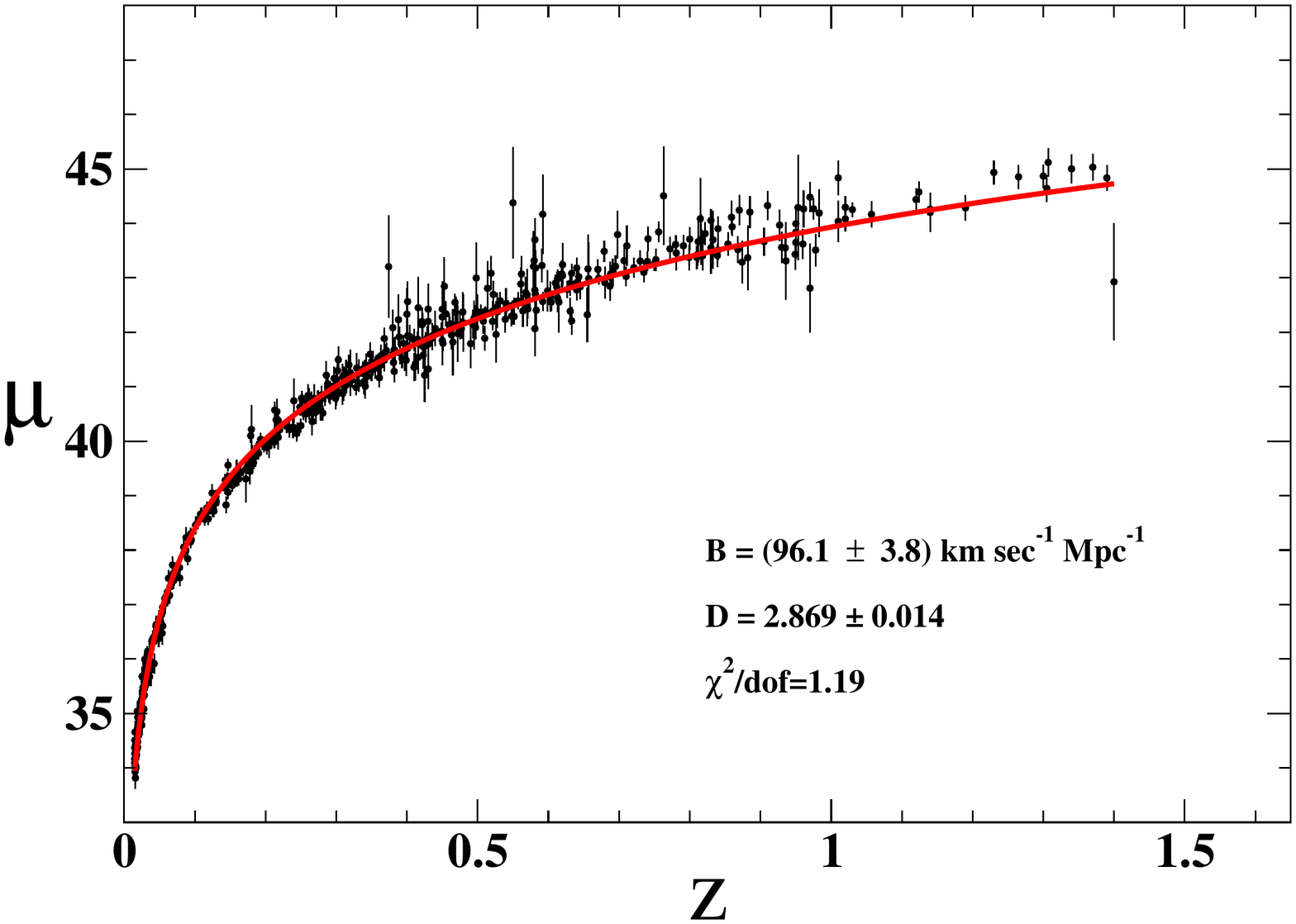}
\caption{Luminosity distance for a pure fractal universe. Data from
  UNION2 catalog.}
\label{fig:SnIa}
\end{figure}

In order to compare this model with the observations of the supernova
Ia data, let us consider the well-known formula of the luminosity distance/redshift relation
 for on-center LTB models \cite{Cosmai:2013iga} (see also \cite{Nogueira:2013ypa}
 for similar interesting approaches). Recently
\cite{Fanizza} within a purely inhomogeneous and anisotropic
framework, an exact geometrical expression for the angular distance
has been evaluated by solving the Sachs equation. {As shown in
  \cite{Fanizza:2014baa}, this solution reduces to the well-known
  formula for the angular distance of on-center LTB models by
  performing a coordinate transformation. Hence, thanks to the Etherington relation,
  we relate the angular and the luminosity distance as both function of redshift $z$
  and} we use the usual expression, given by:
\begin{equation}
d_L =(1+z)^2 A(t(z), r(z))
\end{equation}
where
\begin{align}
\frac{dt}{dz}&=-\frac{A'(t(z),r(z))}{(1+z)\dot A'(t(z),r(z))}\nonumber\\
\frac{dr}{dz}&=\frac{1}{(1+z)\dot A'(t(z),r(z))}.
\label{eq:geodesic}
\end{align}
Let us notice that for $z=0$ we get $d_L(0)=A(t(0),r(0))=A_0(r)=r$, thanks to the partial redefinition of coordinate discussed before.
Eqs. \eqref{eq:geodesic} come from the geodesic null condition $ds^2=0$ for a photon traveling towards the center of the coordinates system, where the affine parameter is labeled by the redshift (see \cite{Romano:2009xw} for the generalization of this procedure when $k(r)\ne 0$). Given the high non-linearity of the solution \eqref{eq:solution}, Eqs. \eqref{eq:geodesic} can be solved only numerically up to larger redshift. However this is already enough for the purposes of this work.
This is important to stress that in a pure fractal distribution, the
center is established only after the integration over the sphere.
Hence, following our approach, the expression for off-center observer
\cite{Cosmai:2013iga,Fanizza:2014baa} has not to be considered.
Let us then consider the distance modulus:
\begin{equation}
\mu(z)=5\log_{10}\left[ \frac{d_L(z)}{1\,\text{Mpc}} \right]+25
\end{equation}
that is immediately comparable with the experimental data, UNION2 data set,  that consist
of redshift-magnitude of 557 supernovae Ia in which we have for each
supernova the observed distance modulus. We do likelihood analysis
with $B$ and $D$ free parameters in order to find the
best fit values. We do the comparison between the observed
$\mu_{obs}(z_i) \pm \Delta \mu(z_i)$ and theoretical distance modulus
$\mu_{th}$ by performing a standard $\chi^2$ analysis with
\begin{equation}
\chi^2= \sum_{i=1}^{557} \left[ \frac {\mu_{obs}(z_i) - \mu_{th}(z_i, \Delta H, r_0, \Delta)} {\Delta \mu(z_i)} \right],
\end{equation}
we find the best-fit values by requiring the minimization of the
$\chi^2$. The minimization has been obtained using MINUIT package from
CERLIB.  The results are shown in Fig.\ref{fig:SnIa}.
The red curve corresponds to the best fit values for our model
i.e. $D=2.87\pm0.02$ and $B=96\pm\pm 4\text{km} \,
\text{sec}^{-1} \, \text{Mpc}^{-1}$ \footnote{In order to make
  independent the dimension of $B$ by $D$, we considered the radial
  coordinate as measured with respect 1 Mpc, namely
  $r=\frac{r}{1\text{ Mpc}}.$} with a
$\chi^2/\text{dof}=1.19$. Therefore, by introducing a fractal exponent
significantly different from 3, we are able to reproduce the supernova
data without referring to any dark energy. Furthermore, our results
are compatible with no transition to homogeneity.

\begin{figure}[h!]
\centering
\includegraphics[width=8.7cm]{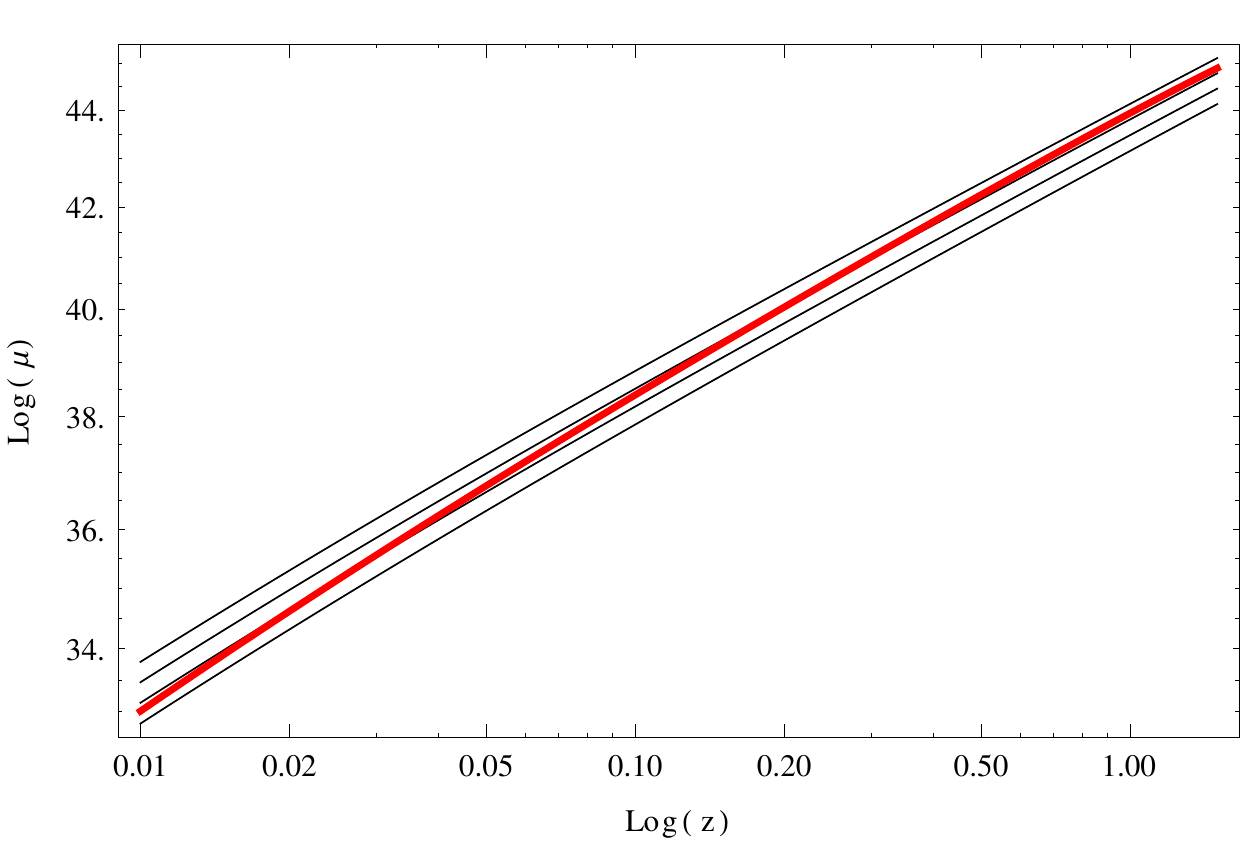}
\caption{Log-Log plot for the distance modulus in term of the
  redshift. Thick red curve indicates the fractal inhomogeneous
  model. Thin black curves refer to several pure matter FLRW models
  with different $H_0$ taken at particular values of $H_0( r)$, from
  $r=1$ Mpc (bottom) to $r=10000$ Mpc (top).}
\label{fig:superposition}
\end{figure}

Let us underline that our solution is decelerated so the acceleration
of the universe is only an apparent homogeneous effect. In fact, as
shown in Fig.\ref{fig:superposition}, our solution (red curve) is the
superposition of different homogeneous FLRW - CDM models with different
$H_0$ (black curves); in particular, at low redshifts, the curve can
be view as a FLRW model with $H_0=H_0( 1\,\text{Mpc})$ but, at high
redshifts, the LTB curve is well described by a FLRW model with
$H_0=H_0( 10000\,\text{Mpc})$.

\begin{figure}[ht!]
\centering
\includegraphics[scale=0.6]{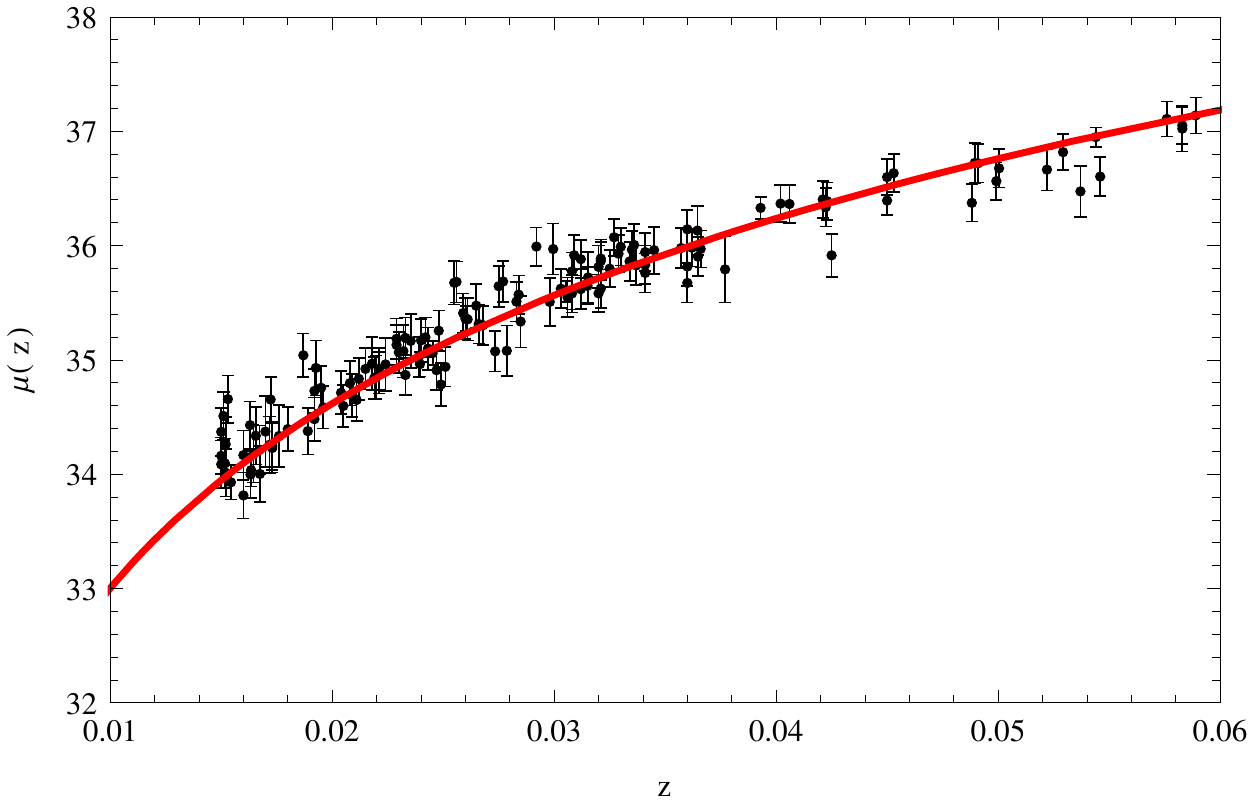}
\includegraphics[scale=0.6]{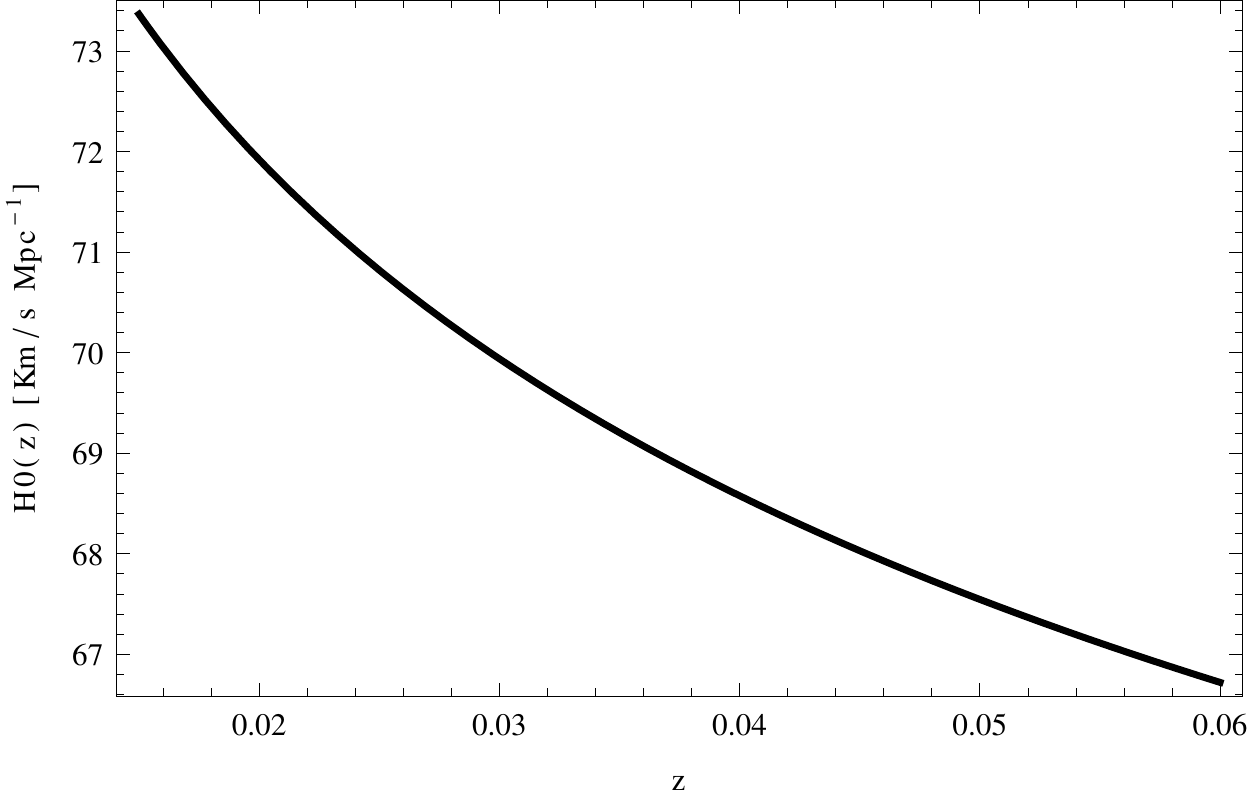}
\caption{Distance modulus and Hubble function in term of redshift at low $z$.}
\label{fig:lowz}
\end{figure}

The validity of our study has to be understood in this sense: the actual
observations about a fractal distribution of matter involve structures up to a
$\sim 100$ Mpc. In this regime, all the cosmological distances
are degenerate so a distribution of matter as considered in
Eq. \eqref{eq:MatterContent} can safely mimic the observed distribution.
Then the application of our result in this range for SNe IA leads to
Fig. \ref{fig:lowz}. For higher redshift, the relation between distances becomes
non-linear and the implementation of a possible fractal behavior on
larger scales may be done more rigourosly on the light-cone rather than on constant-time
hypersurfaces. However the latter extrapolation, as presented in Fig. \ref{fig:SnIa}, just assumes 
that the strongly inhomogeneous behavior extends on scales which are larger than the observed
ones {\it today}. We then use the luminosity distance-redshift relation as an
indirect probe of the possible fractal behavior of matter on largest scales.

According to what we have said up to here, one concern has still to be
addressed i.e. the un-physical divergence of $H_0$.
We note that, in our model, $H_0$ is finite at low redshift
  ($0.01<z<0.05$).  Indeed, let us refer to Fig.~\ref{fig:lowz},
where we considered the lowest redshift data from the UNION2
catalog. As we can see, $H_0$ changes its value from $72$ to $67$ Km/s
Mpc$^{-1}$ within the low redshift range. 
Hence we assume that up to a few Mpc the Hubble law is quiet and that
  at smaller scale our simple description breaks down. Nevertheless,
  we find a very good agreement with the supernovae data even at the
  smallest scale of the UNION2 catalog, as shown in
  Fig.\ref{fig:lowz}.

In summary we have provided an analytical solution for a fractal
distribution of matter in the framework of a LTB model. This
task is not trivial at all and has been achieved by properly take care
about statistical equivalence of the observer's position.  Indeed,
this is really different from the homogeneity of distribution of
matter, which consists of a stronger hypothesis, even if people
usually don't realize it.  Moreover, we used this simplified, but
observationally supported, scenario to show that it is possible to
explain the apparent acceleration of the universe by means of radial
inhomogeneities without introducing dark energy. Nevertheless we
  found that a simple and observationally motivated description
  of large scale galaxy inhomogeneities can describe the Supernovae
  data as well as other exactly inhomogeneous models.
Furthermore  by  using LTB model to compute the
  magnitude-redshift relation
we have found that the best fit to the supernovae data corresponds to
a fractal dimension $D=2.87$ at large scales $r > 100$ Mpc/h, which
is in good agreement with galaxy data \cite{FSL_2011}.  A recent work \cite{ruffini} with a different oversimple inhomogeneous fractal cosmological model the authors
obtain an exponent D=3.36 by means a best fit for UNION 2 supernovae data.

In addition,
we stress that our description is not in contradiction with the
Copernican Principle, as the center point of the LTB model can be
chosen to be in any galaxy, i.e. in any local peak of the conditional
density. A more refined analysis, which will take the CMB Planck data into account
will be presented in a forthcoming work.
\\
\section*{Acknowledgments}
G.F. is supported by Consolidator Grant of the European Research Council (ERC-2015-CoG grant 680886) and by INFN under the program TAsP (Theoretical Astroparticle Physics).

%
%
%

\end{document}